\documentclass[12pt]{iopart}
 \expandafter\let\csname equation*\endcsname\relax
\expandafter\let\csname endequation*\endcsname\relax

\usepackage{url}
\usepackage{amsmath}
\usepackage{graphicx} 
\usepackage{subfig}
\usepackage{iopams}
\usepackage{multirow}
\usepackage{blindtext}
\usepackage{epstopdf} 
\usepackage{xcolor} 
\begin{document}
\title[proton-proton SPS using PFM method]{Study of proton-proton Scattering using Phase Function Method}
\author{Shikha Awasthi$^1$, Anil Khachi$^2$ and O.S.K.S. Sastri$^1$}
\address{$^1$Department of Physics $\&$ Astronomical Sciences\\~~Central University of Himachal Pradesh, Dharamshala 176215, India. \\
~~Email: sastri.osks@hpcu.ac.in\\
$^2$ Department of Physics, St. Bedes College, Himachal Pradesh, India}
\vspace{10pt}
\begin{indented}
\item[]March 2024
\end{indented}
\begin{abstract}\\
\noindent{\textbf{Background:} The study of $np$ and $pp$ scattering, central to understanding nuclear force, still remains an optional topic in many undergraduate nuclear physics curriculum.\\
\textbf{Purpose:} The main thrust of this paper is to study $pp$ scattering using phase function method to obtain the observed S-wave phase shifts and cross section at various energies.\\
\textbf{Methods:} The $pp$ interaction has been modeled by choosing Malfliet-Tjon potential for nuclear part along with screened Coulomb potential. The phase equation has been solved to obtain scattering phase shifts using fourth order RK-method (RK-4).\\
\textbf{Results:} The interaction potential obtained from optimised parameters matches well with the realistic Argonne V18 potential for $^1S_0$ state of $pp$ scattering and the scattering phase shifts as well as the cross section for energies ranging from $1-350$~MeV have been found to be in good agreement with expected data.\\
\textbf{Conclusion:} Introducing phase function method for S-wave ($\ell$=0) could bring this interesting study of nucleon-nucleon scattering to the undergraduate classroom.}
\end{abstract}
\vspace{2pc}
\section{Introduction}
Study of scattering in very light nuclei such as nucleon-nucleon (N-N), nucleon-nucleus (N-n) and nucleus-nucleus (n-n), is an important way to obtain understanding of nuclear force and gain insight into nuclear structure. Nuclear force is hypothesized to be charge independent. That is, the interaction strength involving neutron-proton and neutron-neutron is similar or equivalent to that of proton-proton, excluding its Coulomb repulsion. Hence, it is important to explore the topic of neutron-proton ($np$) and proton-proton ($pp$) scattering through the observed expected phase shift data phase shifts and cross sections at various energies to confirm or refute the hypothesis. The scattering phase shift (SPS) is a vital quantity to understand the nucleon-nucleon scattering. It signifies the changes in phase of the scattering wave occuring as a result of the interaction betwen the nucleons. Proper estimation of the scattering phase shift is crucial to compare experimental results with theoretical predictions and to understand the true nature of nuclear forces.\\
Typically, $np$ scattering is discussed even today, by modeling the interaction to be constant over the inter-nuclear distance as in square well potential \cite{Krane}, even though Yukawa \cite{Yukawa} got Noble prize for his interaction model over 80 years ago. The reason being that it is easy to solve time independent Schr$\ddot{\text{o}}$dinger equation (TISE) analytically for square well as opposed to lack of one, even though very involved, for the Yukawa potential till recently. Although Yukawa potential plays a fundamental role in explaining the strong nuclear force, it is unable to describe the tensor and spin-orbit components of the nuclear force. Because at low energies, the attractive component of nuclear force dominates \cite{Krane} and the Yukawa potential provides an adequate description of nuclear interactions.\\
But at higher laboratory energies, the repulsive core; which results from the strong interaction at extremely short distances; becomes more significant due to the Pauli exclusion principle. Also, at higher laboratory energies, the interacting particles have more kinetic energy which allow them to overcome the attractive forces and come into closer proximity with each other\cite{Bertulani}. At such small distance, the repulsive part of the potential becomes very important and plays important role in reduction of scattering cross section at large scattering angles. Therefore it becomes necessary to include both the repulsive and attractive parts of the nuclear force at higher energies, in order to describe nuclear interactions accurately.\\
This can be achieved by considering complex potential models that contain both attractive and repulsive components of the interaction potential. While potential models like the Woods-Saxon potential and realistic nucleon-nucleon potentials derived from Effective Field Theories (EFT) provide valuable insights into the behavior of light nuclear systems, it's important to note that their applicability and accuracy may be limited when extended to heavier nuclei \cite{Bertulani}. This can be due to the difference in structural properties, shell effects, effective interactions, and the limitations imposed by available experimental data as compared to heavier nuclei. Therefore, specialized models and approaches may be necessary to accurately describe the behavior of light nuclei.\\
There are many realistic nucleon-nucleon (NN) potentials available which are capable of properly describing the interaction between the nucleons \cite{Nijman, Argon}. But in order acquire such an accuracy, they need to adjust large number of parameters (up-to nearly 40 parameters), to calculate the scattering phase shift accurately. It is quite challenging for an undergraduate or graduate student to calculate such a large number of parameters in order to calculate the scattering phase shift. Malfliet-Tjon potential \cite{Malfiet} on the other hand is a three parameter potential, which is based on a non-relativistic approach to study nucleon-nucleon scattering, and provides a comprehensive description of nucleon-nucleon interactions, including both central and tensor components.\\
Recently, we have considered both Yukawa and Malfliet-Tjon potentials to explain the $np$ interaction \cite{nparxiv} for laboratory energies up-to 350 MeV. We observed in our work \cite{nparxiv} on $np$ scattering, that although Yukawa potential is a very good theoretical model used to describe the nuclear force between nucleons for low energy scattering. But at large value of incident projectile energy i.e above 50 MeV the Yukawa potential is not able to predict phase shifts for neutron-proton scattering. Malfliet-Tjon potential on the other hand is a modified Yukawa potential with repulsive term and is able to explain the interaction properly above 50 MeV as well.\\
Most often the scattering phase shifts are obtained from the wave function \cite{Bertulani, Wong, Arya, Bohr} using R-matrix method \cite{rmatrix}, S-matrix method \cite{smatrix}, Jost function method \cite{jost} etc. The cross section and phase shift calculations present in textbooks (if any) are only by considering wave function approach only. Even the formulae to obtain scattering phase shifts given in standard text book by Krane \cite{Krane} is far beyond the scope of under graduate and post graduate courses. So, it is important to provide a simple procedure to calculate scattering phase shifts which will be helpful for under graduate and post graduate students.\\
We have alternatively preferred to use much simpler method known as phase function method (PFM) \cite{Babikov, Calogero}, wherein second order time independent Schr$\ddot{\text{o}}$dinger equation for wave function is cast into a first order Ricatti equation for phase shifts, called phase equation. The interaction potential and scattering energy in center of mass frame are given as input to this phase equation. It is solved numerically to obtain scattering phase shifts which can in turn be utilised to determine scattering cross sections.\\
Proton-proton (pp) scattering which is the scattering of identical nucleons is also an essential topic in nuclear and particle physics course. Although the proton-proton scattering is mentioned in some standard text books \cite{Krane, Wong, Arya, Bohr} at under graduate and post graduate level. However, barely any of the books describe the methods and procedure for calculating scattering phase shifts. The formulae given in standard text book \cite{Krane} to calculate SPS is far beyond the scope of under graduate and post graduate course.\\
In current work, we are studying proton-proton $pp$ interaction by considering Malfliet-Tjon (MT) potential \cite{Malfiet} as interaction model. Proton-proton scattering is very important since it is the interaction which involves the interaction of two fundamental forces of nature i.e the electromagnetic and the nuclear forces, at subatomic scale. Studying $pp$ interaction using phase function method is extremely challenging due to the presence of long range Coulomb interaction alongside the short range nuclear interaction \cite{Preston, AtHulthen, AtHulthen2, Arya}. Hence, one replaces the bare Coulomb potential with a screened one \cite{AtHulthen,AtHulthen2}, considering the observation that an isolated charge is typically surrounded by residual particles due to polarisation in actual experiments. Proton-proton scattering has also been studied using different potentials e.g., square-well potential \cite{Krane}; to calculate cross section; and Yukawa potential \cite{Babenko} to calculate scattering parameters and cross section.\\
In this paper, our goal is to study $pp$ scattering by utilising Malfliet-Tjon (MT) potential as model for nuclear interaction and atomic Hulthén potential \cite{AtHulthen, AtHulthen2} as an ansatz for screened Coulomb interaction. Using this combination, we have solved the phase equation numerically to obtain scattering phase shifts (SPS) and cross sections of $pp$ scattering for energies rannging from $1-350$~MeV. The detail of the implementation have been discussed in step by step simulation methodology for pedagogic purpose.
\section{Methodology}
Methodology is necessary to model the physical system under consideration and to simulate the problem \cite{Aditi1}, and it provides a step-by-step procedure for obtaining the solution. 
The simulation methodology \cite{Aditi1} can be divided into four steps described below:
\begin{enumerate}
\item \textit{Modelling the physics problem:} This involves description of objects, interactions and process followed by formulation of a mathematical model, which typically consists of a dynamical equation with interaction embedded into it and specification of various initial and boundary conditions \cite{AJPaditi, AditiSW}. 
\item \textit{Preparation of the system for numerical solution:} Here, one rephrases the mathematical model into an appropriate choice of units emicable for numerical implementation and sets up the revised iterative equations based on the chosen numerical technique.
\item \textit{Implementation of the numerical method in a computer:} By starting with an algorithm, one has to write a code in a chosen software and optimise the algorithm parameters to obtain a working code for simulation.
\item \textit{Simulation and Discussion of Results:} The system is studied by varying its physical parameters and one discusses their effect in understanding the underlying physics of the problem.
\end{enumerate}
Now, we detail the above four steps for studying proton-proton scattering. 
\subsection{Modeling the physical system}
\subsubsection{\textbf{Description stage:}}
The type of objects involved in a scattering system in the laboratory consists of an incoming projectile, that is accelerated to a certain energy, made to collide with a target that is at rest. In this case, our system is composed of a proton beam, accelerated to lab energies in the range of 1-350 MeV, to be incident on to a target consisting of liquid Hydrogen which can be thought of as collection of protons (mass $m_p$, spin $1/2$). One can neglect the interaction due to electrons which are typically too light to contribute.\\
The interactions between the incoming protons and protons at rest are of two types: 
\begin{enumerate}
\item Short range attraction due to nuclear force and
\item Long range repulsion due to Coulomb force  
\end{enumerate}
Even though, originally Yukawa potential was successful in explaining the nuclear force at low energies, later it was realised that an extra repulsive term similar to  that of Yukawa needs to be added to account for observations at higher energies.\\ 
The process of scattering is more easily understood theoretically, by transofrming the two body system in lab frame to one that consists of a single particle in center of mass system.
The transformation involves position vectors of two particles $m_1$ and $m_2$ with $(\vec{r_1}, \vec{r_2})$ to be related to $(\vec{r}, \vec{R})$, where the first co-ordinate is the relative distance between the two particles and $\vec{R}$ is the center-of-mass (CoM) coordinate. This reduces the two body problem to a single particle system with a mass given by
\begin{equation}
    \mu = \frac{m_{1}m_{2}}{m_{1}+m_{2}} = \frac{m_p}{2}
\end{equation} 
called the reduced mass and its position vector is $\vec{r}$ w.r.t to CoM co-ordinate, chosen as the origin. Since, both nuclear and Coulomb forces are only dependent on the relative distance co-ordinate, they both are of central character and spherical polar co-ordinates are a good choice for reference system.\\
Typically the dynamics, at the microscopic level, are governed by time dependent Schrodinger equation and the state of the system is given by it's wave function: $\Psi(\vec{r,t})$. But, in this work, we are utilising phase function method where in Schrodinger equation is replaced by Ricatti equation in terms of phase shifts. Hence, scattering phase shifts $\delta(r)$ can be considered to be describing state of the system.
\subsubsection{\textbf{Formulation stage:}}
Mathematical Model for Interactions:\\
The Yukawa model \cite{Yukawa}, which introduced the idea of meson exchange, gave an early theoretical foundation for understanding the strong nuclear force. The short-range nature of the nuclear force is effectively explained by the Yukawa model. \\
The inclusion of repulsive interaction of the core along with the attractive part is necessary in order to take into account observed scattering phase shifts at higher energies. Such a potential is proposed by Malfliet and Tjon (MT \cite{Malfiet}, given by:
\begin{equation}
V_N(r) = V_{MT}(r) = -V_A\Big(\frac{e^{-\mu_A r}}{r}\Big) + V_R\Big(\frac{e^{-\mu_R r}}{r}\Big)
\label{MT}
\end{equation}
where, $\mu_R$ is taken equal to $2\mu_A$ in units of $fm^{-1}$. Thus, the MT potential is composed of three parameters containing attractive and repulsive parts.\\
For the charged systems like $pp$ scattering, the Coulomb potential is added to the nuclear interaction potential. But when the particles interact through the Coulomb potential, they do not behave as free particles \cite{AtHulthen}. Therefore, for the charged particle scattering, the conventional scattering theory needs to be revised. Also due to the presence of Coulomb forces, the asymptotic condition which is the basis of traditional scattering theory, is not valid and hence the conventional definitions of scattering phase shift are invalid \cite{Taylor}. Therfore, within the formalismof non-relativistic scattering theory the standard Coulomb potential is replaced by a short range potential such that at large values of distance $r$ the potential should decrease rapidly \cite{Semon}.\\
The ambiguity of long range Coulomb potential can be avoided by considering a screened potential which behave as Coulomb potential at short distances and dies down exponentially at large distances \cite{Preston, AtHulthen}, say for $r$ $>$ $R$. The screened Coulomb potential is commonly utilised in numerous fields of physics like nuclear physics, atomic physics, plasma physics and in many dynamic calculations \cite{AtHulthen}.\\
In this work, we have replaced Coulomb interaction due to protons by Hulthén potential, which is a screened potential \cite{AtHulthen} and is a modified form of Yukawa potential containing exponential term. The Hulthén potential behaves like a Coulomb potential for small values of distance $r$, and compared to the Coulomb potential, it rapidly declines exponentially at large values of $r$; given by
\begin{equation}
V_C(r) = V_H(r) = V_0 \frac{e^{-r/a}}{1-e^{-r/a}} 
\end{equation}
Here $V_0$ is the strength of the potential and  parameter $a$ is chosen in such a way that $aV_0=2k \eta$. Here $\eta$ is a constant quantity known as Sommerfeld parameter given by
\begin{equation}
\eta= \frac{\alpha}{\hbar v}
\end{equation}
Where, $\alpha=Z_1Z_2e^2$ and $v=\sqrt{\frac{2E}{\mu}}$ is the relative velocity of reacting particles at large separation.\\
Hence, $\eta$ will finally be given by
\begin{equation}
\eta=\frac{Z_1Z_2e^2 \mu}{\hbar^2 k}
\end{equation}
Therefore, $aV_0$ after multiplying and dividing by $c^2$ will be equal to
\begin{equation}
aV_0=2k \eta = \frac{2 Z_1Z_2e^2 \mu c^2}{\hbar^2 c^2}~~~fm^{-1}
\end{equation}
Which is a fixed value for a particular interaction, since $\hbar c = 197.327$~MeV-fm, $e^2 = 1.44$~MeV-fm and $\mu c^2$ is the reduced mass of the system in the units MeV/$c^2$. \\
Hence, the total interaction potential for proton-proton (\textit{pp}) system is:
\begin{equation}
        V(r)=-V_A\Big(\frac{e^{-\mu_A r}}{r}\Big)+ V_R\Big(\frac{e^{-\mu_R r}}{r}\Big) + \frac{V_0}{a} \frac{e^{-r/a}}{1-e^{-r/a}} 
        \label{MTH}
    \end{equation}
\textbf{Phase equation for scattering phase shifts:}\\
Phase equation was originally given by Morse \cite{Morse2} and has been briefly discussed here for the sake of completeness. The radial time independent Schr$\ddot{\text{o}}$dinger wave equation (TISE), for partial wave $\ell = 0$ (S-state) is given by
\begin{equation}
\frac{d^2u_{0}(r)}{dr^2}+\frac{2 \mu}{\hbar^2}\big[E-V(r)\big]u_{0}(r) = 0
\label{rTISE} 
\end{equation}
Dividing Eq. \ref{rTISE} by $u_0(r)$, we get
\begin{equation}
\frac{1}{u_0(r)}\frac{d^2u_0(r)}{dr^2} + \frac{2\mu}{\hbar^2}(E-V(r)) = 0 
\label{psi}    
\end{equation}
The wavefunction must satisfy condition $u_0(0) = 0$ at r = 0. Further, at a distance $r_f$ beyond which V(r) is zero, one obtains the asymptotic wavefunction $u_a(r)$, as
\begin{equation}
u_a(r) = \alpha(r)~\sin(kr + \delta(r))
\label{ua}
\end{equation} 
where 
$k = \sqrt{\frac{2\mu E}{\hbar^2}}$. $\alpha(r)$ and $\delta(r)$ represents amplitude and phase in the wavefunction \cite{Bertulani} respectively; depends on distance $r$ and carries information about the region inside the interaction potential $V(r)$ \cite{Viterbo}.\\
Also, the wavefunction and its derivative both need to be continuous at a distance $r_f$. That is, choosing $u_a(r)$ to be asymptotic solution of Eq. \ref{rTISE} for $r > r_f$, we must have
\begin{equation}
u_0(r)\big|_{r = r_f} = u_a(r)\big|_{r = r_f} 
\label{wfc}
\end{equation}
similarly 
\begin{equation}
\frac{du_0(r)}{dr}\bigg|_{r = r_f} = \frac{du_a(r)}{dr}\bigg|_{r = r_f} 
\label{wfdc}
\end{equation}
These two conditions are combined together into a single equation by considering their logarithmic derivatives to satisfy boundary condition \cite{Bertulani, impedance}, obtained as 
\begin{equation}
\frac{1}{u_0(r)}\frac{du_0(r)}{dr}\bigg|_{r = r_f} = \frac{1}{u_a(r)}\frac{du_a(r)}{dr}\bigg|_{r = r_f} 
\label{logder}
\end{equation}
Let us define the logarithmic derivative of wavefunction, based on equation Eq. \ref{logder}, as A(r), equals to
\begin{equation}
A(r) = \frac{1}{u_0(r)}\frac{du_0(r)}{dr} = \frac{1}{u_a(r)}\frac{du_a(r)}{dr}
\label{logdergen}
\end{equation}
Now differentiating $A(r)$ by using first part of Eq. \ref{logdergen}, i.e within the potential region, one obtains
\begin{equation}
\frac{dA(r)}{dr}=\frac{d}{dr}\bigg(\frac{1}{u_0(r)}\frac{du_0(r)}{dr}\bigg) = \frac{1}{u_0(r)}\frac{d^2u_0(r)}{dr^2} - \frac{1}{u_0^2(r)}\bigg(\frac{du_0(r)}{dr}\bigg)^2
\label{Ar}
\end{equation}
From which, we obtain 
\begin{equation}
\frac{1}{u_0(r)}\frac{d^2u_0(r)}{dr^2} = \frac{dA}{dr} + A^2(r) 
\label{Ar2}
\end{equation}
Putting Eq. \ref{Ar2} in Eq. \ref{psi}, we get
\begin{equation}
\frac{dA(r)}{dr} + A^2(r) = \frac{2\mu}{\hbar^2}\big(V(r)-E\big)
\label{Aeqn}
\end{equation}
Eq. \ref{Aeqn} is the Riccati type non linear differential equation which can be linearized to get
radial time independent Schr$\ddot{\text{o}}$dinger equation (TISE) back.\\
Now substituting Eq. \ref{ua} in Eq. \ref{logdergen} by considering asymptotic wavefunction $u_a(r)$, we obtain
\begin{equation}
A(r) = \frac{1}{\alpha ~\sin(kr + \delta)}\frac{d}{dr}[\alpha ~\sin(kr + \delta)]
\label{Acot}
\end{equation}
which on simplification gives
\begin{equation}
A(r) = k~\cot(kr + \delta)
\label{Acot2}
\end{equation}
It is important to keep in mind that while using asymptotic solution, $\delta$ is a constant quantity that needs to be determined at distance $r_f$.\\
Now representing A(r) within the potential well; where the phase shift changes with interaction potential; in terms of wavefunction $u_0(r)$ and taking derivative of Eq. \ref{Acot2}, we get
\begin{equation}
\frac{dA(r)}{dr} = -\frac{(k^2+k\frac{d\delta}{dr})}{\sin^2(kr+\delta)}
\label{Ader}
\end{equation}
In this case, $\delta$ is not treated as a constant but rather as a function of $r$. Further substituting Eq. \ref{Ader} in Eq. \ref{Aeqn}, we obtain
\begin{equation}
-\frac{(k^2+k \frac{d\delta}{dr})}{\sin^2(kr+\delta)} + k^2\frac{\cos^2(kr+\delta)}{\sin^2(kr+\delta)}=\frac{2\mu}{\hbar^2}(V(r)-E)
\end{equation}
Using relation $\cos^2\theta=1-\sin^2\theta$, it gets simplified to result in following phase equation 
\begin{equation}
\boxed{
\frac{d\delta(r)}{dr} = -\frac{2\mu}{\hbar^2}\frac{V(r)}{k}~\sin^2(kr+\delta(r))}
\label{pheqn}
\end{equation}
Equation Eq. \ref{pheqn} is known as phase equation and the function $\delta(r)$, is known as \textit{``phase function"}. Phase function is a function of interaction potential $V(r)$ at a particular interaction distance $r$ and also at a particular value of laboratory energy in terms of wave number $k$. The phase function $\delta(r)$ at any given location $r=R$ yields the value of \textit{``phase shift"} corresponding to interaction potential $V(r)$. A similar expression can be found in work by Morse and Viterbo \cite{Morse2, Viterbo}. Being a nonlinear equation, Eq. \ref{pheqn} has no analytical solutions and hence one has to take the approach of numerical methods.
\subsection{Preparation of the system for numerical solution}
\begin{enumerate}
\item \textbf{Rephrasing the problem in appropriate units}\\
\textit{\underline{Choice of units:}} In nuclear physics, the correct choice of units for energy and distance are MeV and fm respectively. So,  $\hbar c$ gets converted from J-m to $197.329$~MeV-fm.
Reduced mass is chosen in MeV/c$^2$. Hence, phase equation on the right is multiplied and divided by c$^2$ to obtain
\begin{equation}
\boxed{
\frac{d\delta(r)}{dr} = -\frac{2\mu c^2}{\hbar^2c^2}\frac{V(r)}{k_a}~\sin^2(k_ar+\delta(r))}
\label{pheqn1}
\end{equation}
with $k_a = \sqrt{\frac{2\mu c^2 E}{\hbar^2c^2}}$.\\
The physical system which is to be solved analytically or numerically should be expressed in appropriate units. If all the parameters are not in appropriate units, there will be round off errors in the solution obtained. Therefore the physical problem should be rephrased in suitable set of units in order to obtain acceptable solutions.\\
Observing phase equation Eq. \ref{pheqn1}, it is clear that $k$ must have inverse units of $r$, so that argument inside sine function remains dimensionless. We know $\hbar$ has units of J-s and there is no time appearing in the equation. So, idea is to multiply $\hbar$ by speed of light $c$ with units of $m/s$ so that we have energy-meter units. The suitable choice of units in nuclear physics are MeV for energy and $fermi$ $(fm)$ for distance. Hence by considering value of $\hbar c$ equal to $197.329$~MeV-fm, J-m will be converted to MeV-fm. The trick is to multiply and divide factor $\frac{\hbar^2}{2\mu}$ by $c^2$ and choose value of reduced mass in MeV/$c^2$. So, factor will have units of MeV-$fm^2$. This results in $k$ having  
\begin{equation}
k = \sqrt{\frac{E }{\hbar^2 / 2 \mu}} ;~~~~~~\bigg(\text{units~inside~square~root}:~\frac{(MeV)}{(MeV-fm^2)}\bigg)
\end{equation}
units of $fm^{-1}$ as needed. Finally, observe that LHS has units of $fm^{-1}$ and factor multiplying $\sin^2$ term also has same units by choosing potential to have units of MeV. 
This requires, attractive and repulsive strengths, $V_A$ and $V_R$ in Malfliet-Tjon potential to be in units of MeV-fm. Of course, $\mu_A$ has to be in units of $fm^{-1}$. \\
\textit{\underline{Region of Interest (RoI):}} 
One has to understand the nature of interaction to chose appropriate of region of interest in which the phase equation can be solved. Since, nuclear force is of short range, it is important to study the long range Coulomb for various screening radii to decide on the final integration distance, say $r_f$, where Coulomb potential approaches to zero (i.e $V_C$ $\sim$ $10^{-3}$~MeV). We have plotted the screened Coulomb potentials for various values of screening parameter $a$ in Fig.\ref{screened}. 
\begin{figure}[h!]
    \centering
    \includegraphics[width=3in, height=2.6in]{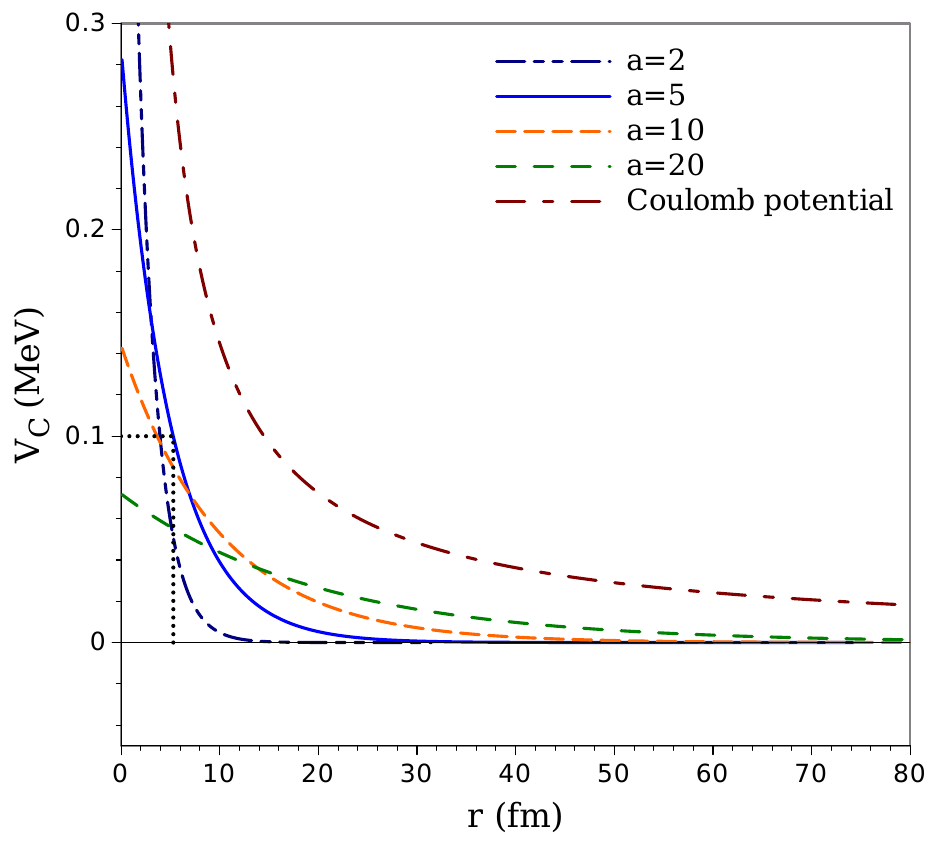}
   \caption{Variation of screened Coulomb potential w.r.t distance `$r$' for different values of screening parameter `a' (fm). The actual Coulomb potential is also plotted for comparison.}
    \label{screened}
\end{figure}
One can observe that with increasing screening parameter value $a$, the potential falls off slowly with distance but has smaller repulsive strength at $r$ close to 0. The cutoff values for $a = 2, 5, 10$ and $20$ are $r_f = 15, 30, 50$ and $90$ respectively. So, RoI would have to be accordingly chosen and sampled with a certain step size, say $h$.\\
It is clear from the Fig.\ref{screened} that the screened Coulomb potential plotted with value of screening parameter $a=5$ following the same trend as that of the variation of original Coulomb potential for proton-proton interaction. Also, the distance up-to which effect of potential is felt for a particular interaction is limited to a certain region. Hence final distance ($r_f$) considered for calculations should be little greater than the total interaction radius of system under consideration. In current work for proton-proton system, we have taken $r_f = 30$~fm since at this distance, the screened Coulomb potential reduces to zero approximately.
\item \textbf{Choice of Numerical Technique:}\\
The first order phase equation can be solved by using Runge Kutta methods. While both RK-2 and RK-4 methods are stable, they have accuracies given by $O(h^2)$ and $O(h^4)$ respectively. Considering that $pp$ expected scattering phase shift data is available up to three decimal places, it is preferable to utilise RK-4 technique. \\
In addition, when compared to the RK-4 method, the RK-2 method requires significantly more iterations to obtain the required accuracy. This could be due to the fact that the RK-2 method requires only two slopes to accomplish the solution, whereas the RK-4 method requires more slopes, resulting in early convergence of the results \cite{Zhaba}.\\
Considering an initial value of function $\delta(r_{0})$= $\delta_0$, Runge kutta methods involve calculating value of function $\delta(r_{1})$, by using the previous values of the function at $r_i$ (i=0,1,2,...n-1). Runge Kutta fourth order method (RK-4) \cite{rk4} consists of four steps with solution given as:
\begin{equation}
\delta(r_{i+1}) = \delta(r_{i}) + \frac{h}{6}(F_1+2F_2+2F_3+F_4)
\end{equation}
where $h$ is the step size and $F_i's$ represent the slopes of the function $f$ at various points within the interval $[0, h]$. Since, the distance $r_f=r_{n-1}$ is a distance at which potential becomes negligible, the corresponding $\delta(r_f)$ will reach a constant value.
\end{enumerate}
\subsection{Implementation of the numerical method in a computer}
The algorithm for obtaining scattering phase shifts using RK-4 method is given in supplemental material provided separately. Once algorithm is carefully tested in a worksheet environment such as Gnumeric/Excel, one can translate it into a code or program in any software such as Scilab, Matlab or Python. \\
Mean square error (MSE) for obtained SPS, is calculated as
\begin{equation}
MSE= \frac{1}{N}\sum_{i=1}^{N}(|\delta^e_i-\delta^o_i|)^2
\end{equation}
Where $\delta^e_i$ and $\delta^o_i$ are the expected phase shifts and obtained phase-shifts for different values of energies $E_i$.\\ 
An optimisation routine based on Variational Monte-Carlo (VMC) tehnique\cite{VMC, Anil, Swapna} is utilised, where in the model parameters are randomly varied in each iteration, so that MSE value converges on to a minimum value.\\
Of course one could use other cost functions such as mean absolute percentage error (MAPE) or chi-squared error ($\chi^2$). The optimisation code is explained briefly in supplemental material provided separately.
\subsection{Experimental observable quantities}
\textbf{\textit{\underline{Cross section}}}\\
After obtaining the SPS $\delta_{\ell}(E)$, one can calculate the partial cross section $\sigma_{\ell}(E)$, which for S-wave ($\ell=0$) is given by:
\begin{equation}
\sigma_{0}(E) = \frac{4\pi}{k_a^2} \sin^2({\delta_{0}(E)})
\label{crossec}
\end{equation} 
\textbf{\textit{\underline{Scattering Parameters}}}\\
Scattering parameters i.e scattering length `$a_0$' and effective range `$r_0$', obtained from scattering phase shifts, can be calculated for low energy $pp$ scattering by using effective range theory \cite{Krane} given by:
\begin{equation}
k \cot(\delta_0)= -\frac{1}{a_0}+ \frac{1}{2}r_0 k^2
\end{equation}
The slope and intercept from straight line plot of $\frac{1}{2}k^2$ vs $k \cot(\delta_0)$ gives scattering parameters `$a_0$' and `$r_0$'.
\section{Results and Discussion}
\begin{figure}[htp]%
    \centering
    \subfloat{{\includegraphics[height = 7cm,width=8.5cm]{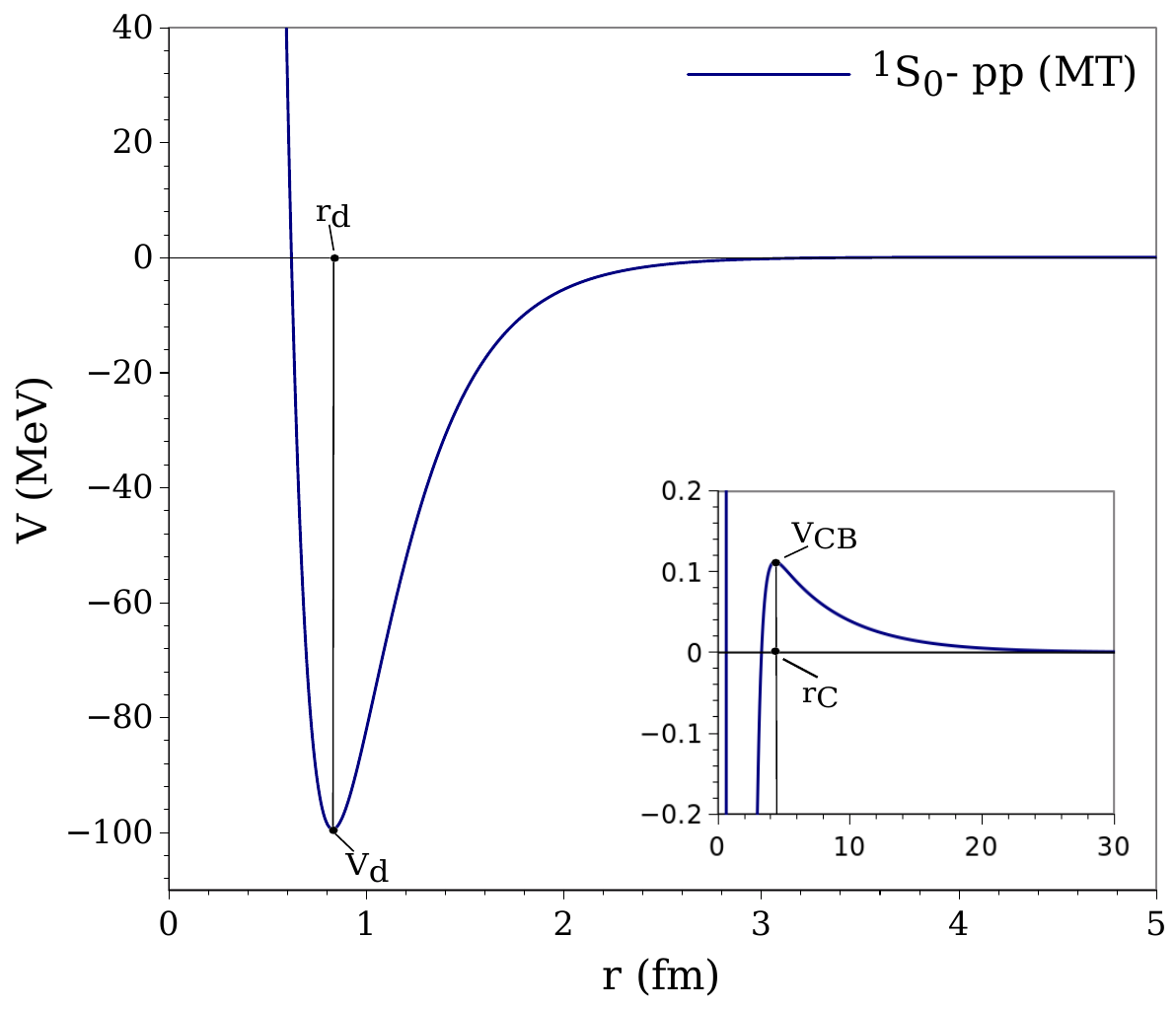} }}%
    \subfloat{{\includegraphics[height = 6.98cm, width=8cm]{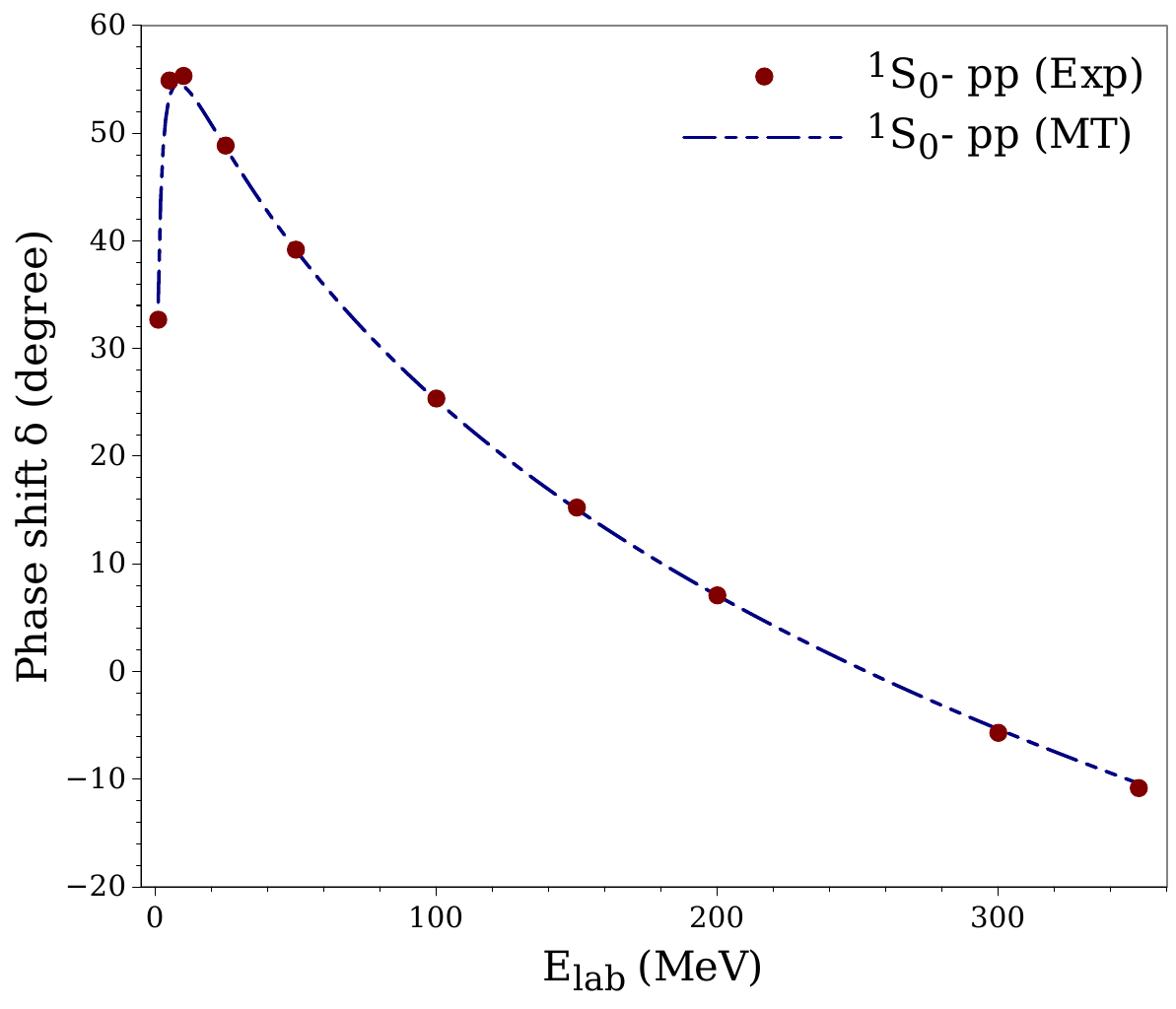} }}%
    \caption{(a) MT interaction potential for $^1S_0$ state of $pp$ scattering using parameters given in Table \ref{parameters} $\textit(left)$. Our potential is matching closely with realistic Argonne AV-$18$ potential given by S. Ohkubo \cite{av182}, inset shows that the interaction potential becomes negligible near $30$~fm and (b) Simulated and expected scattering phase shifts $\textit(right)$ for projectile energies ranging from $1-350$~MeV.}
    \label{pp}%
\end{figure}

\begin{figure}[h!]
    \centering
    \includegraphics[width=3.3in, height=2.9in]{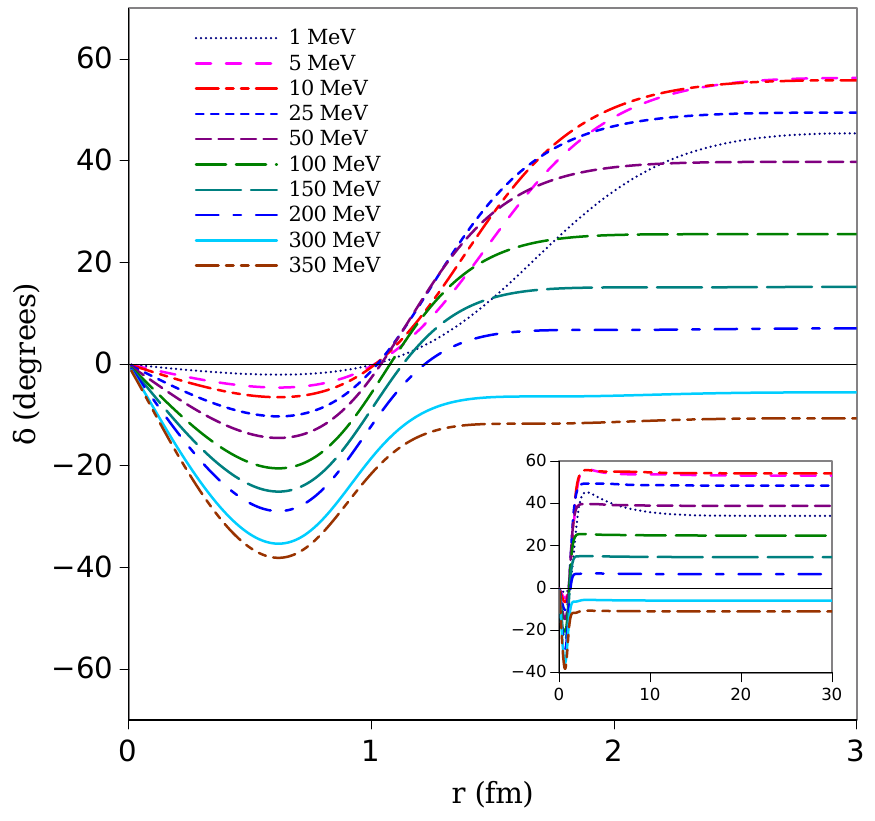}
    \caption{Scattering phase shifts $\delta$ for $^1S_0$ states of $pp$ scattering w.r.t distance $r$ for lab energy $1, 5, 10, 25, 50, 100, 200, 300$ and $350$~MeV. Inset figure shows that the phase shift start becoming constant when the interaction potential starts approaching to zero above $10$~fm and near $30$~fm the phase shift becomes constant.}
    \label{spsr}
\end{figure}

In order to calculate the scattering phase shifts, we have integrated phase equation Eq. \ref{pheqn1} from origin to asymptotic region by using RK-4 method. Mean square error (MSE) is then optimised by using Variational Monte Carlo technique w.r.t the expected phase shift values of \cite{Granada}. The advantage of considering PFM is its simplicity that only interaction potential is required for calculating scattering phase shifts.\\
Model parameters for net interaction potential were determined by minimizing the mean square error (MSE) for different values of screening parameter $a$. The value of $a$ for which mean square error is minimum is finalised. After finalising the value of $a$, the corresponding final integration distance $r_f$ is taken as the distance greater than the distance $r_C$ where Coulomb potential ($V_C$) $\approx$ $10^{-3}$~MeV i.e $r_f >r_C$.\\
For $pp$ interaction, we have obtained screened Coulomb potential for different values of screening parameter $`a'$, e.g $a=2,5,10,20$~fm etc. And for each value of $a$, the depth of potential and corresponding mean square error for simulated scattering phase shifts is calculated. Out of all the value of $a$, the mean square error comes out to be least for $a=5$~fm i.e mean square error= $0.29$. The depth of the interaction potential which matches with that of realistic potential for $^1S_0$ state of $pp$ interaction ($\approx 100$~MeV) also corresponds to value of $a=5$~fm. For $a=5$~fm, the Coulomb potential approaches to zero near $r_C=29$~fm. Hence, we conclude that `a'$=5$~fm is the appropriate value for calculating screened Coulomb potential for $pp$ scattering with minimum mean square error = $ 0.29$ and with final interaction distance $r_f=30$~fm.\\
The optimised model parameters obtained by VMC technique, for MT potential with $a=5$ and $r_f=30$ fm are given in Table \ref{parameters}. We have plotted the final interaction potential for $^1S_0$ state of $pp$ scattering by using the model parameters for MT potential given in Table \ref{parameters} and is shown in Fig. \ref{pp}(a).\\
\begin{table}[h!]
\centering
\caption{Model parameters for MT potential and mean square error (MSE) w.r.t expected data for singlet ($^1S_0$) state of \textit{pp} interaction.}
\begin{tabular}{p{3.3cm}p{3.3cm}p{2cm}p{2cm}|p{2cm}}
\hline \hline
~~~$V_r$ & ~~~$V_a$ & ~~$\mu_A$ & ~$a$ & MSE\\
\small{(MeV-fm)}& \small{(MeV-fm)} & ($fm^{-1}$) & (fm) & \\
\hline
7171.2823  &  1570.9068 & 2.45  & ~~5 & 0.29\\
\hline \hline
\end{tabular}
\label{parameters}
\end{table}
The MT potential obtained by using our optimised parameters is in good agreement with that of realistic Argonne AV-$18$ potential \cite{Argon, av18} given by S. Ohkubo \cite{av182} for $^1S_0$ state of proton-proton interaction thereby validating our approach. It is found that the interaction potential so obtained with parameters has depth $V_d=-99.44$~MeV with equilibrium distance $r_d=0.85$~fm, closely match with that of realistic Argonne AV-$18$ potential \cite{Argon, av182, av18} having depth $\approx$ -103.5 MeV at equilibrium distance $r_d=0.89$~fm. Hence we can validate that the MT potential parameters which we have obtained for $pp$ scattering, describes the interaction well.\\
The Coulomb barrier height $V_{CB}$ is also calculated for interaction potential obtained. $V_{CB}$ is found to be $=0.11$~MeV at distance $r_C=4.3$~fm. The variation beyond matches with screened Coulomb potential for $a=5$~fm. This is because nuclear force being short range, becomes $\sim$ 0 for distance of about $4-5$~fm itself.\\
Simulated phase shifts obtained by considering MT potential parameters from Table \ref{parameters} along with expected phase shifts for $^1S_0$ state of \textit{pp} scattering are given in Table \ref{nppp1}. It is evident that that all the phase shifts have discrepency less than $6$ ($\%$). The variation of scattering phase shifts for lab energies $1-350$~MeV are shown in Fig. \ref{pp}(b) along with expected data points from Granada \cite{Granada}. From the figure, one can observe an excellent match at all energies.\\
\begin{table}[htp!]
\centering
\caption{Scattering phase shifts (in degrees) for $^1S_0$ states of $pp$ interaction, w.r.t expected data \cite{Granada}, for lab energies up-to $350$~MeV. The absolute percentage error is also given for each data point in fourth column and the mean absolute percentage error(MAPE) is given at the bottom.}
\begin{tabular}{l c c c c}
\hline \hline
Lab energy &  $^1S_0$(Expected)\cite{Granada} &  $^1S_0$ (Malfliet-Tjon) & Absolute percentage\\
$E_{lab_{i}}$ (MeV) &  $\delta_{e_i}$ &  $\delta_{o_i}$ & error ($\%$)\\
\hline \hline
1 &  32.677  & \ \  34.285 & 4.92\\
5 &  54.895 & \ \  53.348 & 2.82\\
10 & 55.32  & \ \  54.383  & 1.69\\
25 &  48.848 & \ \  48.662 &  0.38\\
50 &  39.182 & \ \  39.137 &  0.11\\
100 &  25.357 & \ \  25.199 & 0.62\\
150 &  15.229 & \ \   15.074  & 1.02\\
200 &  7.076 & \ \  7.057  & 0.27\\
300 &  -5.694 & \ \  -5.353  & 5.99\\
350 &  -10.828 & \ \  -10.397  & 3.98\\
\hline \hline
MAPE ($\%$) &&& 2.18 \\
\hline \hline
\end{tabular}
\label{nppp1}
\end{table}
One of the main advantage of phase function method is that, it provides the information of how phase shifts progressively builds-up due to interaction potential which can be observed by plotting phase shift w.r.t interaction distance $r$. This feature helps students to understand the effect of interaction potential on phase shift as the interaction distance increases. We have plotted phase shifts w.r.t distance for different laboratory energies. The phase shifts $\delta(r)$ w.r.t distance $r$ for lab energies $1, 5, 10, 25, 50, 100, 200, 300$ and $350$~MeV are plotted in Fig. \ref{spsr} by considering the parameters from Table \ref{parameters}. It is clear from the figure that the phase shifts remains negative up-to approximately $1$~fm, till the repulsive core of the potential and then starts increasing towards positive side when the potential becomes attarctive. Phase shift reaches to a constant value at $r=r_f=30$~fm where the potential becomes negligible. The phase shift values vary for each value of incident energy as well.\\
Partial cross section ($\sigma_0$) for $^1S_0$ state of \textit{pp} scattering is calculated by using Eq. \ref{crossec} and is given in figure \ref{crossec2}. Since we have calculated cross section corresponding to $\ell=0$ partial wave (i.e S-state) only, therefore there is a slight deviation of the calculated $\sigma_0$ from experimental data taken from tables (\textit{Table 4, 5}) published by R. A. Arndt $et.al.$ in their work \cite{Arndtcrossec}.\\
\begin{figure}[h!]
    \centering
    \subfloat{{\includegraphics[height = 7.3 cm,width=9cm]{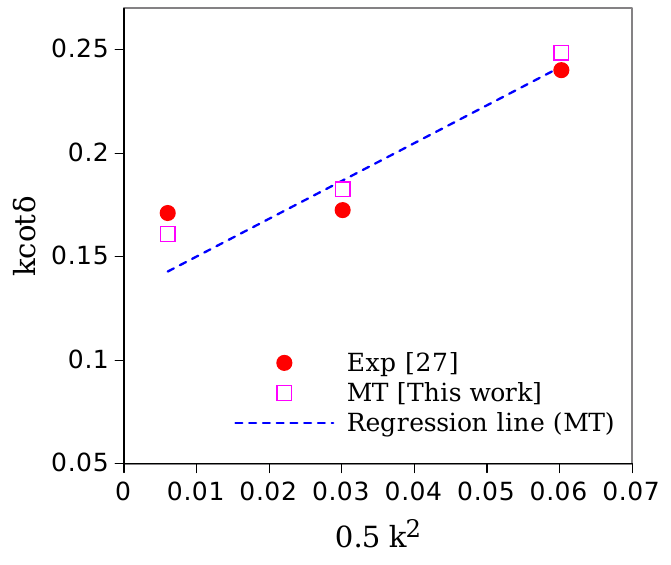} }}%
    \caption{ Plot of $k\cot(\delta)$ vs 1/2 $k^2$ plot for scattering phase shifts of $^1S_0$ state of \textit{pp} scattering corresponding to different lab energies up-to $10$~MeV.}
    \label{crossec2}
\end{figure}
\begin{figure}[h!]
    \centering
    \includegraphics[height = 7.3cm, width=8.5cm]{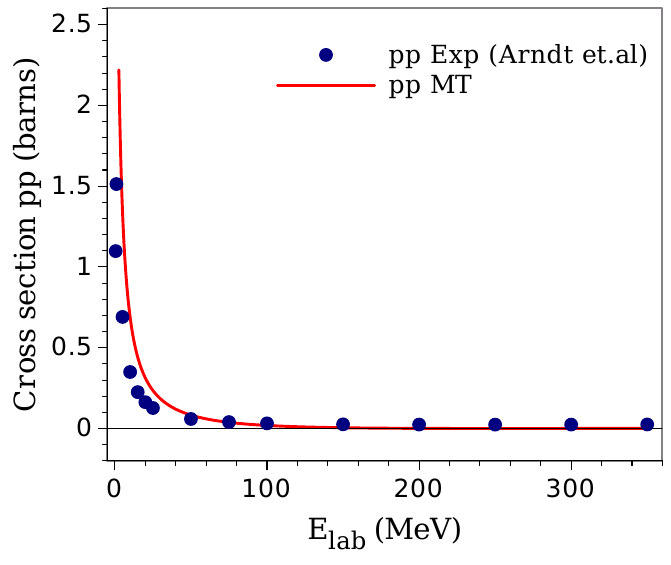}
    \caption{Calculated and experimental (\textit{data taken from table published by R. A. Arndt $et.al.$ in their work} \cite{Arndtcrossec}) cross section ($\sigma_0$) for $^1S_0$ state of \textit{pp} scattering}
    \label{crossec2}
\end{figure}
\begin{table}[h!]
\centering
\caption{Scattering length `$a$' and effective range `$r_0$' parameters calculated for $^1S_0$ state of \textit{pp} scattering along with the expected values, the errors are given within brackets}
\begin{tabular}{l|c c c c}
\hline \hline
States & $a(fm)$ \cite{Miller}  & $a(fm)$ (calc.) & $r_0(fm)$ \cite{Miller}  & $r_0(fm)$ (calc.)\\
\hline	
$^1S_0$-$pp$ & -7.8063 ($\pm$ 0.003) & -7.5857 ($\pm$ 0.0008)& \ \ 2.794 ($\pm$ 0.001) & 1.8272 ($\pm$ 0.021)\\
\hline \hline
\end{tabular}
\label{scat}
\end{table}
The Scattering parameters i.e scattering length `$a_0$' and effective range `$r_0$' parameters for $^1S_0$ state of \textit{pp} scattering are also calculated. These scattering parameters have been calculated from slope and intercept of $k cot(\delta_0)$ $vs$ $(0.5 \times k^2)$ plots given in Fig. \ref{crossec2}(a). The obtained scattering parameters along with expected data are tabulated in Table \ref{scat}. It can be seen that the calculated scattering parameters are in good agreement with the results obtained by Miller $et.al.$\\
Based on the above discussions, we conclude that:
\begin{enumerate}
\item The scattering phase shifts calculated using simple phase function method along with our VMC optimisation procedure shows very good results with mean square error of 0.29 w.r.t expected data of Granda group \cite{Granada}. 
\item Following the pedagogical approach, the algorithm as a pseudo code for our VMC optimisation procedure and the implementation in Gnumeric worksheet to obtain scattering phase shift has been given in the supplemental material provided separately for easy access of students.
\item The interaction potential which we have obtained by using our optimisation procedure for $^1S_0$ state of \textit{pp} scattering matches well with the realistic Argonne Av-$18$ potential given by Wiringa $et.al$\cite{av18, av182}, thereby validating our approach.
\item The scattering phase shifts calculated using simple phase function method by using only four parameters of total interaction potential, three for Malfliet-Tjon potential and one for the screened Coulomb potential, unlike most of the relistic potentials which require approximately 40 parameters for the calculation.
\item The scattering parameters, i.e scattering length `$a$' and effective range `$r_0$' calculated by using obtained scattering phase shifts also matches with the expected values given by Miller $et.al$ \cite{Miller}.
\item Scattering cross section so calculated shows matching trend with the experimental cross section data for $pp$ scattering taken from \cite{Arndtcrossec}. Discrepencies in the calculated cross section are due to the fact that in this current work, we have not considered higher partial waves for $pp$ scattering which is our future work.
\end{enumerate}
\section{Conclusion}
The proton-proton ($pp$) interaction modeled, using phenomenological Malfliet-Tjon potential for nuclear interaction and screened Coulomb potential has been found to explain the observed scattering phase shifts for the $^1S_0$ singlet state. The obtained scattering phase shifts match the expected data with a mean absolute percentage error of 2.18 $\%$ and mean square error of $0.29$. The scattering cross-sections and scattering parameters are found to be reasonably close to the respective expected data. This procedure based on phase function method for understanding $np$ and $pp$ scattering can be easily integrated into classroom and laboratories at the undergraduate level. Further, students can be encouraged to take-up projects by considering other interactions like nucleon-nucleus ($nd$, $pd$, $n\alpha$, $p\alpha$) and nucleus-nucleus ($\alpha-\alpha$) scattering systems.\\ \\ 
\textbf{Declarations:}\\ \\
Funding: We hereby state that there is no funding received for this work.\\ \\
Conflict of Interest: There is no conflict of interest whatsoever.
\clearpage
\noindent{\textbf{References}}\\

\end{document}